\documentclass[
     final            
  ]
  {aipproc}

\layoutstyle{6x9}

\begin{document}

\title{Extraction of nucleon resonances from global analysis of meson production reactions at EBAC}


\author{Hiroyuki~Kamano}{
  address={Excited Baryon Analysis Center, Thomas Jefferson National Accelerator Facility, Newport News, Virginia 23606, USA}
}

\begin{abstract}
We report the current status of exploring the dynamical aspect of 
the excited nucleon states through the comprehensive coupled-channels analysis of meson production reactions 
at the Excited Baryon Analysis Center of Jefferson Lab.
\end{abstract}

\maketitle

\paragraph{Introduction}
The meson production reactions on the nucleon target are
a powerful tool to study the excited nucleon ($N^\ast$) states.
Increasing amount of the precise data from electron/photon beam facilities, 
such as JLab, Mainz, Bonn, GRAAL, and SPring-8,
has opened a great opportunity
for establishing the spectrum of high $N^\ast$ states at $W > 1.6 $ GeV
and extracting various form factors associated with the $N^\ast$ states.

The $N^*$ states, however, couple strongly to the meson-baryon continuum states and 
appear only as resonance states in the reaction processes. 
Such strong couplings will affect significantly the $N^*$ properties and 
cannot be neglected in extracting the $N^*$ parameters from 
the data and making physical interpretations.
It is thus well recognized nowadays that a comprehensive and consistent study of 
meson production reactions with 
$\pi N$, $\eta N$, $\pi\pi N$, $KY$, $\omega N$, ... final states 
based on a coupled-channels framework
is crucial for a reliable extraction of the $N^\ast$ properties.

In this contribution, we present the status of exploring the dynamical aspect of 
the $N^\ast$ states through a comprehensive analysis
of $\pi N$, $\gamma N$, and $N(e,e')$ reactions,
which is being conducted at the Excited Baryon Analysis Center (EBAC) of Jefferson Lab.
The analysis is based on an dynamical coupled-channels model,
the EBAC-DCC model (see Ref.~\cite{msl} for the details),
within which the multi-channel unitarity including the 3-body $\pi\pi N$
channel is fully taken into account.

\paragraph{EBAC-DCC analysis (2006-2009)}
We summarize the status of the DCC analysis 
of meson production reactions at EBAC in Table~\ref{tab:history}.
During the developing stage of EBAC (middle column of Table~\ref{tab:history}),  
hadronic and electromagnetic parameters of the EBAC-DCC model were determined by analyzing 
$\pi N\rightarrow \pi N$~\cite{jlms07} and $\pi N \rightarrow \eta N$~\cite{djlss08}
up to $W=2$ GeV, $\gamma N \rightarrow \pi N$~\cite{jlmss08} up to $W=1.6$ GeV, 
and $N(e,e'\pi)N$~\cite{jklmss09} up to $W=1.6$ GeV and $Q^2=1.5$ GeV$^2$. 
The model constructed by these analysis is then applied to $\pi N\to \pi\pi N$~\cite{kjlms09}
($\gamma N\to\pi\pi N$~\cite{kjlms09-2}) and gave reasonable predictions for the invariant mass distributions
in the wide energy region from threshold up to $W=2$ GeV ($W=1.5$ GeV),
indicating the important role of the coupled-channels framework in achieving a consistent and unified description
of various meson production reactions.
\begin{table}
\begin{tabular}{lcc}
\hline
&\tablehead{1}{c}{b}{2006 -- 2009} &\tablehead{1}{c}{b}{2010 --} \\
\hline
Number of &5 channels& 7 channels \\
coupled channels & ($\pi N,\eta N, \pi\Delta, \sigma N, \rho N$)& 
($\pi N,\eta N, \pi\Delta, \sigma N, \rho N$, $K\Lambda$, $K\Sigma$) \\
\\
$\pi N \to \pi N$            & $W < 2.0$ GeV~\cite{jlms07} & $W < 2.1$ GeV \\
$\gamma N \to \pi N$         & $W < 1.6$ GeV~\cite{jlmss08} & $W < 2.0$ GeV \\
$\pi N \to \eta N$           & $W < 2.0$ GeV~\cite{djlss08} & $W < 2.0$ GeV \\
$\gamma N \to \eta N$        &    ---        & $W < 2.0$ GeV \\
$\pi N \to K\Lambda,K\Sigma$ &    ---        & $W < 2.1$ GeV \\
$\gamma N \to K \Lambda$     &    ---        & $W < 2.1$ GeV \\
$e N \to e' \pi N$           & $W < 1.6$ GeV, $Q^2 < 1.5$ GeV$^2$~\cite{jklmss09} & $W < 1.6$ GeV, $Q^2 < 6$ GeV$^2$ \\
\hline
\end{tabular}
\caption{Summary for the dynamical coupled-channels analysis of meson production reactions at EBAC
during 2006-2009 (middle column) and since 2010 (right column).}
\label{tab:history}
\end{table}

\begin{figure}[tb]
\includegraphics[height=0.22\textheight,clip]{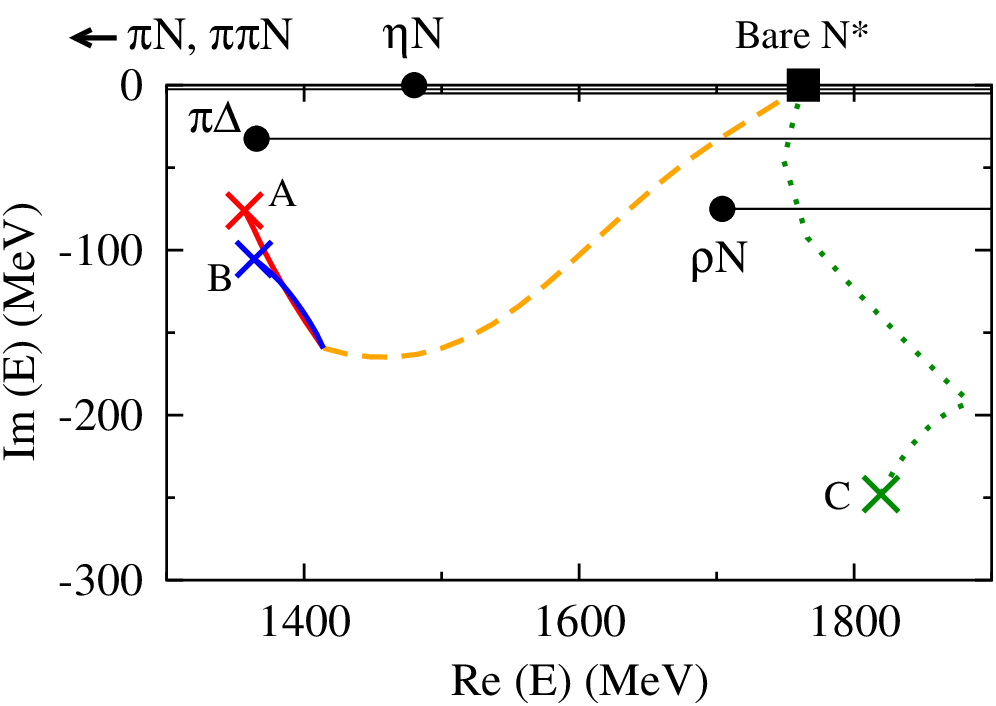}
\caption{
Dynamical origin of $P_{11}$ nucleon resonances in the EBAC-DCC model~\cite{sjklms10}:
Two resonance poles (A and B) around the Roper energy and the higher resonance pole (C) corresponding to
$N^\ast(1710)$ in the PDG notation are generated from a single bare state as a result of the dynamics of the multi-channel
reactions.
}
\label{fig:p11}
\end{figure}

\paragraph{Extraction of $N^\ast$}
With the reaction model constructed from the analysis, we have extracted the $N^\ast$ 
spectrum~\cite{sjklms10,knls10}
and the $N$-$N^\ast$ electromagnetic transition form factors~\cite{ssl2}, which are
defined respectively by the pole positions and the residues of the scattering amplitudes in the complex energy plane.
Figure~\ref{fig:p11} shows the spectrum and the dynamical origin of the $P_{11}$ resonances within the EBAC-DCC model,
indicating the crucial role played by the multi-channel reaction mechanisms for 
determining the dynamical origin of the resonance states: 
(a) the two-pole structure of the Roper resonance due to the existence of the $\pi\Delta$ unitarity cut and 
(b) one-to-many correspondence between the bare and physical resonance states~\cite{sjklms10}.
By definition, the bare $N^\ast$ states in our framework can be related with the states obtained from 
the static hadron structure calculations such as quark models in which the coupling to the meson-baryon continuum states is neglected.
The results of Fig.~\ref{fig:p11} imply that the correspondence between the hadronic states obtained from the static hadron structure calculations 
and the physical resonance states is very complicated in general due to the reaction dynamics.

\begin{figure}[tb]
\includegraphics[height=0.18\textheight,clip]{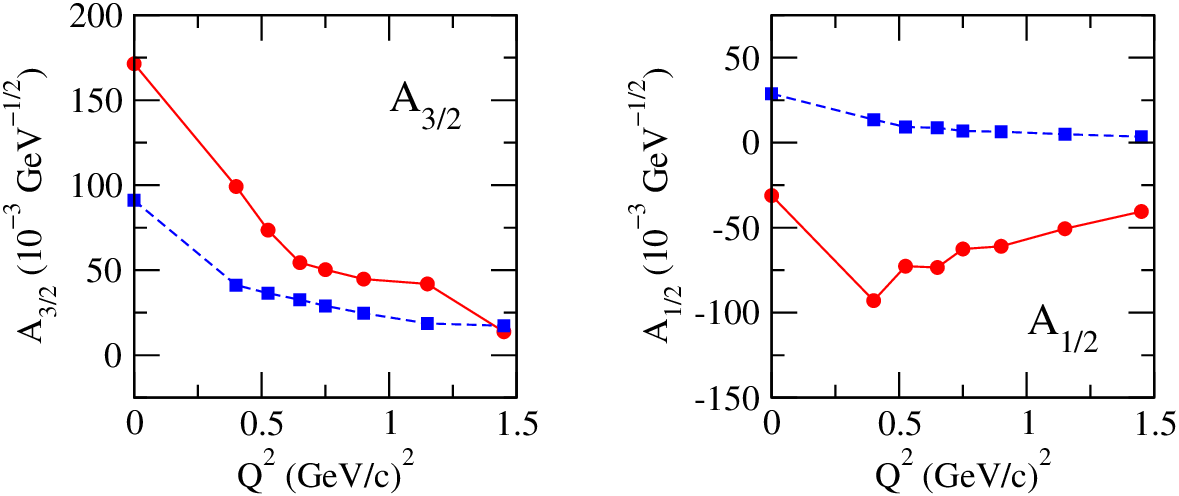}
\caption{
The $Q^2$ dependence of the electromagnetic transition form factors 
between the nucleon and the fist $D_{13}$ state in the EBAC-DCC model~\cite{ssl2}.
Real (imaginary) part of the form factors is displayed as
solid circles connected with solid line (squares connected with dashed line).
}
\label{fig:d13}
\end{figure}

Figure~\ref{fig:d13} shows the electromagnetic transition form factors between the nucleon and 
the first $D_{13}$ state extracted from the EBAC-DCC model.
This is the first successful extraction of the form factors beyond $\Delta (1232)$ within a fully dynamical coupled-channels framework. 
We observe that, as a result of the coupling to the reaction channels, the extracted form factors are complex, 
even though the bare form factors are purely real.

\paragraph{EBAC-DCC analysis (2010 - )}
To make further progress, recently we have extended the EBAC-DCC model by including the $K\Lambda$ and $K\Sigma$ channels 
and started a fully combined analysis of the $\pi N, \gamma N \rightarrow \pi N, \eta N, K\Lambda, K\Sigma$ reactions 
(right column of Table~\ref{tab:history}).

\begin{figure}[tb]
\includegraphics[width=0.95\textwidth,clip]{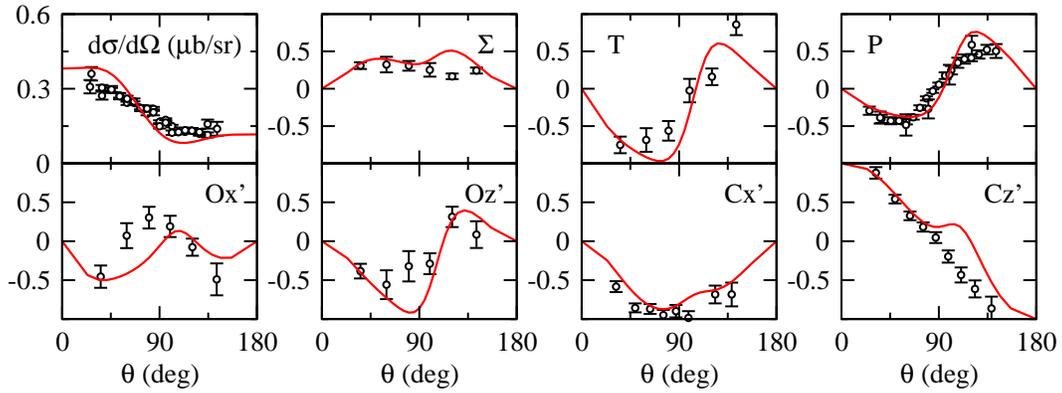}
\caption{
Comparison of our current result (preliminary) with the available data of $\gamma p \to K^+\Lambda$ 
at $E_\gamma = 1421$ MeV ($W=1883$ MeV).
The data are from Refs.~\cite{g1c06,g1c07,g11a10,graal07,graal09}.
}
\label{fig:gkl}
\end{figure}

Figure~\ref{fig:gkl} is the preliminary results of the $\gamma p\to K^+\Lambda$ analysis at 
$E_\gamma = 1421$ MeV.
Although it needs further improvements, our model describes
overall behavior of the available differential cross sections and polarization observables.
Regarding $KY$ photoproduction reactions, it is worthwhile to mention that the so-called ``(over-) complete experiments'' is ongoing at CLAS
for $\gamma p \to K^+ \Lambda$ and $\gamma n \to K^0 \Lambda$ reactions,
in which \emph{all} 16 observables (unpolarized cross sections and 15 polarization observables) are measured.
Those data are expected to be a key to understanding the $N^\ast$ states around $W=2$ GeV, 
which are still not well-established.
A potential impact of the complete experiments on the amplitude extraction
has been investigated extensively in Ref.~\cite{shkl}.

\paragraph{Summary and outlook}
In summary, we have presented the current status of the EBAC-DCC analysis.
Once we construct a new model by a fully combined analysis of 
the $\pi N, \gamma N \rightarrow \pi N, \eta N, K\Lambda, K\Sigma$ reactions, 
we will reinvestigate $N^\ast$ spectrum and examine how the inclusion of $\eta N$ and $KY$ reaction data affects the pole positions.
We also extend our $N$-$N^\ast$ electromagnetic transition form factors to high $Q^2$ up to 6 GeV$^2$
by analyzing all the existing CLAS data of single pion electroproductions.
Also, we will keep improving our model by including other reactions such as $\omega N$ and the double-pion channels, $\pi\pi N$, in the analysis.
Unfortunately, there is essentially no data for the differential cross sections of $\pi N \to \pi \pi N$ in the resonance region.
We hope such data will be provided by hadron beam facilities such as J-PARC.

Finally, as a new direction, we have recently started an application of our dynamical 
coupled-channels approach to meson physics~\cite{3pion}.
As a first step, we examine various issues associated with the three meson decays of 
heavy mesons and meson excited states particularly focusing on the 
effect of the 3-body unitarity on the decay processes,
which, for example, will be relevant to the 
physics program to search for hybrid mesons proposed by GlueX and CLAS12.

\begin{theacknowledgments}
The author would like to thank B.~Juli\'a-D\'{\i}az, T.-S.~H.~Lee, 
A.~Matsuyama, S.~X.~Nakamura, T.~Sato, and N.~Suzuki for their collaborations at EBAC.
This work was supported by the U.S. Department of Energy, Office of Nuclear Physics Division, under 
Contract No. DE-AC05-06OR23177, under which Jefferson Science Associates operates the Jefferson Lab.
\end{theacknowledgments}

\bibliographystyle{aipproc}   

\end{document}